\documentclass[12pt]{article}

\begin{document}
\begin{titlepage}
\begin{center}
\hfill    CERN-PH-TH/2007-217\\

\vskip 1cm

%\hfill HUPD-0109\\
%~{} \hfill hep-ph/0107164\\
%\vskip 1cm

{\large \bf {Four Zero Neutrino Yukawa Textures in the Minimal Seesaw
Framework}}

\vskip 1cm

Gustavo C. Branco$^{a}$,\footnote{gbranco@ist.utl.pt} 
David Emmanuel-Costa$^a$,\footnote{david.costa@ist.utl.pt}
M. N. Rebelo$^{a,b}$  \footnote{Presently at CERN on sabbatical leave 
from IST,  
margarida.rebelo@cern.ch \\ and rebelo@ist.utl.pt}
and  Probir Roy$^{c,d}$ \footnote{DAE Raja Ramanna Fellow, 
presently at SINP, probirr@gmail.com}
\vskip 0.05in

{\em a Departamento de F{\'\i}sica and Centro  de F{\'\i}sica
Te{\'o}rica de Part{\'\i}culas (CFTP),\\
Instituto Superior T\'{e}cnico (IST), Av. Rovisco Pais, 1049-001
Lisboa, Portugal \\
b CERN, Department of Physics, Theory Division, CH-1211 Gen\` eve 23, 
Switzerland \\ 
c Tata Institute of Fundamental Research,  Homi Bhabha Road, 
Mumbai 400 005, India \\
d Saha Institute of Nuclear Physics, Block AF, Sector 1, Kolkata 700
064, India}

\end{center}

\vskip 3cm

\begin{abstract}
We investigate, within the Type I seesaw framework, the
physical implications of zero textures in the Yukawa couplings
which generate the neutrino Dirac mass matrix $m_D$.
It is shown that four is the maximal number of 
texture zeroes compatible with the observed leptonic mixing 
and the assumption that no neutrino mass vanishes. We classify 
all allowed four-zero textures of $m_D$ into two categories
with three classes each. We show that the different classes,
in general, admit CP violation both at low and high energies.
We further present the constraints obtained
for low energy physics in each case.
The r\^ ole of these zero textures in establishing a connection 
between leptogenesis and low energy data is analysed in detail. 
It is shown that it is possible in all cases to completely specify
the parameters relevant for leptogenesis in terms of light
neutrino masses and leptonic mixing together with the 
unknown heavy neutrino masses.
\end{abstract}

\end{titlepage}

\newpage
\section{Introduction}
Recent impressive experimental progress towards determining the masses
and mixing angles of the three known light neutrinos has brought
urgency to the task of unravelling the flavour
structure of the neutrino mass matrix $m_\nu$. It has been pointed out
\cite{Frampton:2002yf} that, in the case of Majorana neutrinos, it is
not possible to completely determine the structure of $m_\nu$ from
feasible experiments. This is one of the motivations for introducing
some theoretical input aimed at reducing the number of free
parameters. One interesting possibility is the imposition of zeroes in
the elements of $m_\nu$ \cite{Frampton:2002yf}. Another is to assume
the vanishing of det $m_\nu$ \cite{Branco:2002ie}.
Several papers have analysed in detail the consequences of imposing
zeroes directly in the elements of $m_\nu$, starting with at least two
zero textures \cite{Frampton:2002yf}, \cite{Dev:2006qe}.  Implications
of single texture zeroes were studied in detail in
\cite{Merle:2006du}. In fact, the existence of vanishing mass matrix
elements may reflect the presence of a family symmetry acting in the
leptonic sector \cite{Grimus:2004hf}, \cite{Kaneko:2007ea}. It is then
more natural to investigate and classify the appearance of zeroes in
the fundamental mass matrix appearing in the Lagrangian rather than in
$m_\nu$ which, at least within the seesaw framework, is a derived
quantity. Therefore we focus our attention 
on the Yukawa couplings which lead to
the neutrino Dirac mass matrix $m_D$, once spontaneous symmetry
breaking occurs. One can then see how  
zeroes in $m_D$ affect $m_\nu$ which we
take to be related to $m_D$ by the Type I seesaw relation.
Throughout, by "texture" we shall refer to a
configuration of $m_D$ containing zeroes in some of its elements.  In
the Froggatt-Nielsen approach \cite{Froggatt:1978nt} texture zeroes
correspond to {\bf extremely suppressed} entries, which can be taken
effectively as zeroes. The stability of zeroes in neutrino mass
matrices under quantum corrections in type I seesaw models has been
studied in Refs. \cite{Hagedorn:2004ba}, \cite{Antusch:2005gp},
\cite{Mei:2005qp}.  One also needs to be aware that renormalisation
group effects can be quite large in the case of quasi-degenerate (inverted
hierarchical) light neutrino masses \cite{Dighe:2006qd},
\cite{Dighe:2007nh}.

In this paper we classify and analyse the physical implications of all
neutrino Yukawa coupling matrices with four zero textures in the {\bf
Weak Basis} (WB) where the charged lepton and the righthanded Majorana
neutrino mass matrices are diagonal and real. For simplicity, we work
within the framework of the Type I seesaw, where three righthanded
singlet neutrinos are added to the Standard Model (SM). 
The case of only two righthanded
heavy neutrinos leads to one zero neutrino mass and in this case only
one zero textures and some of the two zero textures are allowed
experimentally \cite{Ibarra:2003up}, \cite{Guo:2006qa}. 
With three
heavy righthanded neutrinos and the additional requirement that none
of the physical neutrino masses  vanishes, we show
that four is the maximal number of zeroes in textures of $m_D$ that
are compatible with the available data on neutrino mixing.  We
organize all such four zero textures into classes and discuss the
physical implications of each class. The
imposition of texture zeroes in the Yukawa couplings has the advantage
of allowing for the possibility of establishing a connection between
low energy physics and physics at high energies, in particular
leptogenesis \cite{Fukugita:1986hr}. 

It is by now established that new sources of CP violation beyond the
Kobayashi-Maskawa mechanism of the Standard Model (SM) are required in
order to dynamically generate the observed Baryon Asymmetry of the
Universe (BAU) \cite{Gavela:1994dt},  
\cite{Huet:1994jb}, \cite{Anderson:1991zb}, \cite{Buchmuller:1993bq},
\cite{Kajantie:1995kf}, \cite{Fromme:2006cm}.
The scenario of baryogenesis
through leptogenesis has been rendered especially appealing by the
discovery of neutrino oscillations which provides evidence
for nonvanishing neutrino masses.
In general, there is no direct
connection between CP violation at low energies and that entering in
leptogenesis \cite{Branco:2001pq}, \cite{Rebelo:2002wj}.  It has been
shown, however, that such a connection arises in models with
texture zeroes in $m_D$ \cite{Frampton:2002yf}, \cite{Ibarra:2003up}
\cite{Kaneko:2002yp}, \cite{Branco:2002xf}.
This is a question that we analyse in the present work for each one of
the classes of allowed four zero textures and we conclude 
that it is possible in all cases to completely specify
the parameters relevant for leptogenesis in terms of light
neutrino masses and leptonic mixing together with the 
unknown heavy neutrino masses.

Texture zeroes are clearly not WB invariant. For definitness we analyse
the allowed four zero textures in the WB 
in which both the charged lepton mass matrix and the heavy
righthanded Majorana neutrino mass matrix are diagonal, 
as mentioned earlier. The question of how to 
recognise a flavour model corresponding to a given set of texture 
zeroes, when written in a different basis, was addressed in 
Ref~\cite{Branco:2005jr}. It was shown there that some sets of 
texture zeroes imply the vanishing of certain CP-odd WB invariants.
The relevance of CP-odd WB invariants in the analysis of texture 
zero ans\" atze is due to the fact that texture zeroes lead in general
to a decrease in the number of CP violating phases. 
 
This paper is organised as follows. In section 2 we set the notation
and present our framework.
We show in section 3, based on what is already known
experimentally about leptonic mixing and on the condition that no
neutrino mass vanishes, why textures with five or more zeroes 
in $m_D$ are ruled out. In section 4 we enumerate
the possible classes of four zero textures that are allowed 
and give for each one of them  low energy relations leading 
to physical constraints. CP violation and related WB invariants for the 
surviving four zero textures are discussed in section 5.
In section 6 we analyse the physical implications of four zero textures
both for low energies and for leptogenesis. Our 
conclusions are summarised in section 7.

\section{Notation and framework}
We work in the context of the minimal Type I seesaw with three generations of
righthanded neutrinos which are singlets of SU(2). We do not
extend the Higgs sector and therefore we do not include Majorana mass
terms for lefthanded neutrinos. After spontaneous symmetry breaking, 
the leptonic mass terms are given by:
\begin{eqnarray}
{\cal L}_m  &=& -[\overline{{\nu}_{L}^0} m_D \nu_{R}^0 +
\frac{1}{2} \nu_{R}^{0T} C M_R \nu_{R}^0+
\overline{l_L^0} m_l l_R^0] + h. c.  \nonumber \\
&=& - [\frac{1}{2}  n_{L}^{T} C {\cal M}^* n_L +
\overline{l_L^0} m_l l_R^0 ] + h. c.,
\label{lep}
\end{eqnarray}
where $n_L$ is a column vector with $n_L^T = ({\nu}_{L}^0, {(\nu_R^0)}^c)$,
while $M_R$, $m_D$, and $m_l$ respectively denote the  
righthanded neutrino Majorana mass matrix, 
the neutrino Dirac mass matrix and the charged
lepton mass matrix in family space. The superscript 0 signifies the fact 
that the corersponding fields are eigenstates of flavour.  
The matrix $\cal M $ is given by:
\begin{eqnarray}
{\cal M}= \left(\begin{array}{cc}  
 0  & m_D \\
m^T_D & M_R \end{array}\right). \label{calm}
\end{eqnarray}

We assume that the scale of $M_R$
is much higher than the electroweak scale $v \simeq$ 246 GeV. 
Upon diagonalisation of 
the matrix $\cal M$, we are left with ${\cal D} ={\rm diag} (m_1, m_2, m_3$,
$M_1, M_2, M_3)$, containing three light and three heavy Majorana neutrinos. 
The charged current interactions can then be written as
\begin{equation}
{\cal L}_W = - \frac{g}{\sqrt{2}} \left( \overline{l_{iL}} 
\gamma_{\mu} K_{ij} {\nu_j}_L +
\overline{l_{iL}} \gamma_{\mu} G_{ij} {N_j}_L \right) W^{\mu}+h.c.,
\label{phys}
\end{equation}
where  $\nu_j$ and $N_j$ denote the 
light and the heavy neutrinos respectively and $K$ and $G$ are 
3 x 3 blocks of a unitary 6 x 6 matrix that diagonalises
the symmetric matrix ${\cal M}$. 
The light neutrino 
masses and mixing angles are obtained to an excellent approximation from: 
\begin{equation}
U^\dagger m_{eff}  U^\star =d, \label{14}
\label{eq7}
\end{equation}
where $m_{eff} = - m_D {M_R}^{-1} m^T_D \equiv m_\nu$ 
computed in the WB where $m_l$ is diagonal, and $d$ = diag ($m_1, m_2, m_3$). 
The unitary matrix $U$
in Eq.~(\ref{14}) is the Pontecorvo, Maki, Nakagawa
and Sakata (PMNS) matrix \cite{pmns} relating the mass eigenstate neutrinos
$\nu_i \ (i = 1,2 ,3)$ to the flavour eigenstate neutrinos $\nu_f \ (f=
e,\mu,\tau)$ 
by ${\nu_f}_L = U_{fi}{\nu_i}_L$. It coincides with $K$ 
in Eq.~(\ref{phys}) up to corrections of order 
$v^2/M^2$, which we ignore. 
The Yukawa textures that we analyse are imposed in 
the weak basis  where $M_R$ and $m_l$ are real and diagonal. In this WB,
all CP violating phases are contained in $m_D$. From Eq.~(\ref{eq7}) 
and the definition of $m_{eff}$, one can write $m_D$ in the Casas and 
Ibarra parametrisation
\cite{Casas:2001sr}:
\begin{equation}
m_D = i U {\sqrt d} R {\sqrt D},
\label{udr}
\end{equation}
where $D$ stands for $M_R$ in the WB where it is diagonal and $R$ is
a general complex orthogonal matrix.
It is clear, by construction, that mixing and CP violation at low
energies are blind to the matrix $R$. However this last matrix  
is relevant for leptogenesis.

\section{Inadmissibility of more than four zero textures}

We shall now demonstrate that, in the framework specified earlier,
all five zero textures in $m_D$ are ruled out. Let us start with the
general form 
\begin{equation}
m_D = \left[
\begin{array}{ccc}
 a_{1} & a_{2} & a_{3} \\ 
 b_{1} & b_{2} & b_{3} \\ 
 c_{1} & c_{2} & c_{3} \end{array} \right],
%\label{exe}
\end{equation}
Since we are assuming that none of the neutrino masses vanishes
we conclude from the definition of $m_{eff}$ that the determinant 
of $m_D$ cannot be zero. Therefore patterns of $m_D$ with one full
row of zeroes or one full column of zeroes are ruled out, as 
well as patterns with zeroes distributed in a quartet,
i.e. in four elements (ij), ($l$k), (ik), ($l$j) 
where i,j,k,$l$ can be 1, 2 or 3, with $l\neq$i, k$\neq$j.
We are thus left with patterns where the zeroes are placed in such a way
that invariably two rowns and two columns would have two zeroes 
simultaneously. Together with the requirement that there is no quartet 
of zeroes this leads to several different possibilities where, in 
each case, one nonzero entry of $m_D$ is in a row and a column
where all other entries are zeroes. Some examples are
\begin{eqnarray}
\left[ 
\begin{array}{ccc}
 a_{1} & a_{2} & 0 \\ 
 0  & 0 & b_{3} \\ 
 c_{1} & 0 & 0
\end{array}
\right], & \left[ 
\begin{array}{ccc}
 a_{1} & a_{2} & 0 \\ 
 0 & 0 & b_{3} \\ 
 0 & c_{2} & 0
\end{array}
\right],& \left[ 
\begin{array}{ccc}
 0 & a_{2} & 0 \\ 
 0 & b_{2} & b_{3} \\ 
 c_{1} & 0 & 0
\end{array}
\right], \nonumber \\ 
\left[ 
\begin{array}{ccc}
 a_{1} & 0 & 0 \\ 
 0 & b_{2} & 0 \\ 
 0 & c_{2} & c_{3}
\end{array}
\right],  
& \left[ 
\begin{array}{ccc}
 0 & a_{2} & a_{3} \\ 
 b_{1} & 0 & 0 \\ 
 0 & c_{2} & 0
\end{array}
\right],& \left[ 
\begin{array}{ccc}
 0 & a_{2} & a_{3} \\ 
 0 & b_{2} & 0 \\ 
 c_{1} & 0 & 0
\end{array}
\right].
%\label{cat2}
\end{eqnarray}

Since we work in a WB where
$M_R$ is diagonal the resulting matrix $m_{eff}$, 
for the five zero textures under discussion, is always
block diagonal. Furthermore, the fact that we are in a WB 
where the charged lepton mass matrix is also diagonal, implies that 
these textures lead to two-family mixing only, which is already
ruled out experimentally. Indeed, it is already known that there are
two large mixing angles in the PMNS matrix and as a result  
all five zero textures are ruled out. 
 
\section{Four zero textures}
   In this section, we classify all different possible four zero textures
for $m_D$  in a WB where $M_R$ and $m_l$ are real and diagonal 
with no vanishing neutrino mass.
Among patterns of four zero textures in $m_D$, the 
nonvanishing det $m_D$ condition rules out the occurrence of 
three of the zeroes 
in the same row or in the same column, as well as zeroes distributed 
in a quartet, as explained in the previous section. 
Block diagonal patterns such as
\begin{equation}
\left[
\begin{array}{ccc}
a_{1} & 0 & 0 \\ 
0 & b_{2} & b_{3} \\ 
0 & c_{2} & c_{3}
\end{array}
\right],\ \ \left[ 
\begin{array}{ccc}
a_1 & 0 & a_3 \\ 
0 & b_2 & 0 \\ 
c_{1} & 0 & c_{3}
\end{array}
\right],\ \ \left[ 
\begin{array}{ccc}
a_1 & a_2 & 0 \\ 
b_{1} & b_{2} & 0 \\ 
0 & 0 & c_3
\end{array}
\right]\ \   \label{bla}
\end{equation}
lead to two-family mixing only and are therefore ruled out. \\

The allowed remaining patterns can be 
split into two categories:
\bigskip

(i) those with two orthogonal rows;
\bigskip

(ii) those with two orthogonal columns and no pairs of orthogonal rows; \\

The first category can be divided into three classes corresponding to:
\bigskip

(i)(a) orthogonality of the first and second rows, leading to:
\begin{equation}
{m_{eff}}_{12} = {m_{eff}}_{21}=0.
\label{eff12}
\end{equation} 

(i)(b)  orthogonality of the first and third rows, leading to:
\begin{equation}
{m_{eff}}_{13} = {m_{eff}}_{31}=0.
\label{eff13}
\end{equation}

(i)(c)  orthogonality of the second and third rows, leading to:
\begin{equation}
{m_{eff}}_{23} = {m_{eff}}_{32}=0.   
\label{eff23}
\end{equation} 

There are eighteen different cases in (i)(a). Six of them
have two zeroes in the first row and two zeroes in the second row,
as for example:
\begin{equation}
\left[ 
\begin{array}{ccc}
 0  &  0 & a_{3} \\ 
 0  & b_{2} & 0 \\ 
 c_{1} & c_{2} & c_{3}
\end{array}
\right],  \left[ 
\begin{array}{ccc}
 0  &  0 & a_{3} \\ 
b_{1} & 0 & 0 \\ 
c_{1} & c_{2} & c_{3}
\end{array}
\right].
\label{esta}
\end{equation}
Another six different cases have two zeroes in the 
first row, one zero in the second row and one zero in the 
third row as in:
\begin{equation} 
\left[ 
\begin{array}{ccc}
 0  & a_{2} & 0   \\ 
b_{1}  & 0 & b_{3}\\ 
0  & c_{2} & c_{3}
\end{array}
\right], 
\left[ 
\begin{array}{ccc}
 0  & a_{2} & 0 \\ 
 b_{1}  & 0 & b_{3} \\ 
 c_{1} & c_{2} & 0
\end{array}
\right].
\label{outra}
\end{equation}
Finally, six different cases are 
obtained with one zero in the first row, two zeroes in the 
second row and one zero in the third row as in:
\begin{equation}
\left[ 
\begin{array}{ccc}
0  &  a_{2} & a_{3} \\ 
b_{1} & 0 & 0 \\ 
c_{1} & 0 & c_{3}
\end{array}
\right], \left[ 
\begin{array}{ccc}
0  &  a_{2} & a_{3} \\ 
b_{1} & 0 & 0 \\ 
c_{1}  & c_{2} & 0
\end{array}
\right].
\label{pois}
\end{equation}
There are another eighteen different cases in (i)(b). These are
obtained from those in (i)(a) exchanging the second with the third
row. The cases in (i)(c) are also eighteen different ones obtained
from those in (i)(a) by exchanging the first row with the third one.
Each case in category (i) has one symmetric
pair of nondiagonal zero entries in $m_{eff}$. Since $m_{eff}$
is symmetric by construction, due to its Majorana character, 
off diagonal zeroes always come in pairs.

Textures in category (ii) are obtained by transposing those in
category (i) and discarding those already considered in (i). 
There are eighteen cases altogether in category (ii). In all of these, 
two columns are
orthogonal to each other, each having two zeroes, and there is one column 
without zeroes. 
This category can again be divided into three classes:
\bigskip

(ii)(a) six cases with two zeroes in the first row, these cases are 
given explicitly by:. 
\begin{eqnarray}
\left[ 
\begin{array}{ccc}
 0 & 0 & a_{3} \\ 
 0  & b_{2} & b_{3} \\ 
 c_{1} & 0 & c_{3}
\end{array}
\right], & \left[ 
\begin{array}{ccc}
 0  &  0 & a_{3} \\ 
b_{1} & 0 & b_{3} \\ 
 0 & c_{2} & c_{3}
\end{array}
\right],& \left[ 
\begin{array}{ccc}
 0 & a_{2} & 0 \\ 
 0 & b_{2} & b_{3} \\ 
 c_{1}  & c_{2} & 0
\end{array}
\right],  \nonumber \\
\left[ 
\begin{array}{ccc}
 0  & a_{2} & 0 \\ 
 b_{1}  & b_{2} & 0 \\ 
 0 & c_{2} & c_{3}
\end{array}
\right], & \left[ 
\begin{array}{ccc}
a_{1} &  0 & 0 \\ 
b_{1} & b_{2} & 0 \\ 
c_{1} & 0 & c_{3}
\end{array}
\right],& \left[ 
\begin{array}{ccc}
a_{1} & 0 & 0 \\ 
b_{1} & 0 & b_{3} \\ 
c_{1} & c_{2} & 0
\end{array}
\right].
\label{cat2}
\end{eqnarray}
These verify the conditions:
\begin{equation}
\left|{m_{eff}}_{11}{m_{eff}}_{23}\right| = 
\left|{m_{eff}}_{12} {m_{eff}}_{13}\right|, \qquad
\arg(m_{11}m_{23}m^*_{12}m^*_{13}) =0.
\label{iia}
\end{equation}
Note that $\arg(m_{ii}m_{jk}m^*_{ij}m^*_{ik})$ is rephasing invariant.
In Ref. \cite{Branco:2005jr} one example of this class was discussed.

(ii)(b) six cases with two zeroes in the second row,  which are
obtained from the patterns in (ii)(a) by interchanging the first with
the second row. These verify the conditions:
\begin{equation}
\left|{m_{eff}}_{22}{m_{eff}}_{13}\right| = 
\left|{m_{eff}}_{21} {m_{eff}}_{23}\right|, \qquad
\arg(m_{22}m_{13}m^*_{21}m^*_{23}) =0.
\label{iib}
\end{equation}

(ii)(c) six cases with two zeroes in the third row,  which are
obtained from the patterns in (ii)(a) by interchanging the first with
the third row. These verify the conditions:
\begin{equation}
\left|{m_{eff}}_{33}{m_{eff}}_{12}\right| = 
\left|{m_{eff}}_{31} {m_{eff}}_{32}\right|, \qquad
\arg(m_{33}m_{12}m^*_{31}m^*_{32}) =0.
\label{iic}
\end{equation}
Eqs.~(\ref{iia}), (\ref{iib}) and (\ref{iic}) are of the form:
\begin{equation}
{m_{eff}}_{ii}{m_{eff}}_{jk} = {m_{eff}}_{ij} {m_{eff}}_{ik}
\label{frm}
\end{equation}
with $i,j,k$ different from each other and no sum implied.

It can be checked that all allowed cases in category (i) as well as 
in category (ii) contain the same number of independent parameters
in $m_D$ and in all such cases one can rephase away three of the phases.
The counting of independent parameters in $m_{eff}$ is also
the same in all cases, as will be seen in the next section. 
Moreover, we shall analyse in section 6 the implications of Eqs.~(\ref{eff12}),
(\ref{eff13}), (\ref{eff23}), (\ref{iia}), (\ref{iib}), (\ref{iic})
corresponding to the two categories, each one with three different 
classes.

Notice that, although we are considering weak bases with the maximum  
number of zeroes allowed by experiment, 
together with the assumption that no neutrino mass vanishes,
the resulting matrix $m_{eff}$ contains at most one zero 
nondiagonal entry. We are not considering here the possibility of 
fine-tuning between the parameters of $m_D$ and those of $M_R$
leading to additional zeroes due to special cancellations.
This indicates that imposing texture zeros in the WB where $m_l$
and $M_R$ are diagonal does not allow to generate any of the
two zero patterns considered in Ref.~\cite{Frampton:2002yf}. 
It is already known that not all 
of these patterns can realised through seesaw  \cite{Kageyama:2002zw},
\cite{Branco:2007nn}

\section{CP Violation and Weak Basis Invariants}
We start by recalling the general counting of the number
of parameters contained in the lepton mass matrices and
then consider the special case of textures with four zeros
in $m_D$. In the WB where $M_R$ and $m_l$ are diagonal and real,
leptonic mixing and CP violation are encoded in $m_D$,
which is an arbitrary complex $3 \times 3$ matrix. The latter contains
nine real moduli and nine phases. Of these, only six
phases are physical, since three phases can be removed 
by simultaneous rephasing of $\nu_L$, $l_L$. So
$m_D$ is left with nine real moduli plus six  
phases. Taking into account the three eigenvalues of
$M_R$, we have in this WB a total of eighteen  parameters
including six phases. This equals the number of physical parameters
to wit, three light neutrino masses, three heavy neutrino masses, 
plus six mixing angles and six CP violating phases in 
the first three rows of a  $6 \times 6$ complex unitary matrix
\cite{Branco:1986my}
which we have denoted as $K$ and $G$ in Eq.~(\ref{phys}).
It is interesting to notice that  
the number of independent physical phases in Eq.~(\ref{udr}) is also 
six, three in the
PMNS matrix and three required to parametrise the orthogonal complex
matrix $R$.

Textures with four zeros in $m_D$ lead to 
a strong reduction in the number of parameters, since there are only 
five real parameters and two phases after rephasing. This gives rise to
interesting phenomenological implications which are analysed in 
detail in the next section. In particular, it will be shown that 
in all  four zero textures classified by us, the matrix $R$,
which plays an important r\^ ole in leptogenesis, can be fully 
expressed in terms of low energy parameters  
entering in $U$ and $d$. This establishes a direct connection
between leptogenesis and low energy data. Moreover, this link exists both in 
the cases of unflavoured and flavoured leptogenesis.

In scenarios where flavour does not play an important r\^ ole, 
the lepton number asymmetry resulting from the decay of the $N_j$ 
heavy Majorana neutrino is given by \cite{sym}:
\begin{eqnarray}
\varepsilon _{N_{j}}
&=& \frac{g^2}{{M_W}^2} \sum_{k \ne j} \left[
{\rm Im} \left(({m_D}^\dagger m_D)_{jk} ({m_D}^\dagger m_D)_{jk} \right)
\frac{1}{16 \pi} \left(I(x_k)+ \frac{\sqrt{x_k}}{1-x_k} \right)
\right]
\frac{1}{({m_D}^\dagger m_D)_{jj}}   \nonumber \\
&\simeq& \frac{g^2}{{M_W}^2} \sum_{k \ne j} \left[ (M_k)^2
{\rm Im} \left((G^\dagger G)_{jk} (G^\dagger G)_{jk} \right)
\frac{1}{16 \pi} \left(I(x_k)+ \frac{\sqrt{x_k}}{1-x_k} \right)
\right]
\frac{1}{(G^\dagger G)_{jj}}. \nonumber \\
\label{rmy}
\end{eqnarray}
In Eq. (20) $M_k$ are the heavy neutrino masses and we have neglected
terms of order $v^2/M^2_k$.  The variable $x_k$
is defined as  $x_k=M_k^2/M_j^2$ and the function $I(x_k)$ is given by
$I(x_k)=\sqrt{x_k} \left(1+(1+x_k) (\log x_k - \log (1+x_k)) \right)$.
Eq. (20) has been obtained after summing over all charged leptons
$l_i^\pm$ ($i$ = e, $\mu$ , $\tau$).
From the Casas and Ibarra parametrisation we obtain:
\begin{equation}
m^\dagger_D m_D = \sqrt{D} R^\dagger d R
\sqrt{D}.  
\label{drr}
\end{equation}
In this framework, leptogenesis is insensitive to the low energy 
CP violating phases appearing in $U$ and can occur
even without CP violation at low energies \cite{Rebelo:2002wj}.
Actually, leptogenesis depends on other parameters beyond $\varepsilon
_{N_{j}}$ 
and involves thermodynamic processes that have been analysed by 
several authors \cite{Buchmuller:2003gz}, \cite{Buchmuller:2004nz}
\cite{Buchmuller:2004tu}, \cite{Giudice:2003jh}. It was pointed out 
recently that, under certain conditions, flavour matters in 
leptogenesis \cite{Barbieri:1999ma}, 
\cite{Fujihara:2005pv}, \cite{Pilaftsis:2005rv},
\cite{Abada:2006fw}, \cite{Nardi:2006fx}, 
\cite{Abada:2006ea}. In this case we must take into account
the separate lepton $i$ family asymmetry generated from the decay 
of the $k$th heavy Majorana neutrino  which depends on the combinations
\cite{Fujihara:2005pv}
Im$\left( (m_D^\dagger m_D)_{k k^\prime}(m_D^*)_{ik} (m_D)_{ik^\prime}\right) $
and 
Im$\left( (m_D^\dagger m_D)_{k^\prime k}(m_D^*)_{ik} (m_D)_{ik^\prime}\right) $
Clearly, when one works with 
separate flavours, the matrix $U$ does not cancel out and one
is led to the interesting possibility of having viable
leptogenesis even in the case of $R$ being a real matrix \cite{Branco:2006hz},
\cite{Pascoli:2006ie}, \cite{Branco:2006ce}, \cite{Uhlig:2006xf}. 

Next, we show explicitly that four zero textures lead in general
to CP violation both at low and high energies. The strength of leptonic 
CP violation of Dirac-type, which can be observable through neutrino
oscillations, is controlled by the WB CP-odd invariant  \cite{Branco:1986gr}
\begin{equation}
I_1 \equiv
{\rm tr} \left[ h_{eff}, {h_l} \right]^3 = -6i \ \Delta \ 
I_{CP}, \qquad {\rm with} \qquad
I_{CP} \equiv {\rm Im} \ ({h_{eff}}_{12}{h_{eff}}_{31}{h_{eff}}_{23}), 
\label{trr}
\end{equation}
where $h_{eff} = m_{eff} m_{eff}^\dagger$,  
$h_l = m_l m_l^\dagger$ and  
$\Delta = ({m_{\mu}}^2 - {m_e}^2 ) ({m_{\tau}}^2 -{m_e}^2 ) 
({m_{\tau}}^2 -{m_{\mu}}^2 )$.
In order to show that this CP-odd invariant does not vanish,
in spite of the four zeroes in $m_D$, we have to examine the structure of
$h_{eff}$. For definiteness, let us consider the configuration 
\begin{equation}
m_D =
\left[ 
\begin{array}{ccc}
 a_{1} & a_{2} & a_{3} \\ 
 b_{1} & 0 & 0 \\ 
 0 & 0 & c_{3}
\end{array}
\right]
\label{exe}
\end{equation}
belonging to case (c), category (i).  
Three phases can be rephased away so that one is left with only two
phases which can be placed for instance at the entries (1,1) and (1,2). From 
Eq.~(\ref{exe}) and the definition of $m_{eff}$ one obtains the
following structure for the latter:
\begin{equation}
m_{eff} \equiv
\left[ 
\begin{array}{ccc}
 m_{11} &  m_{12} &  m_{13} \\ 
 m_{12} &  m_{22} &  m_{23} \\ 
 m_{13}  & m_{23} &  m_{33}
\end{array}
\right] = \left[ 
\begin{array}{ccc}
 c_{11}  &  c_{12} &  r_{13} \\ 
 c_{12} &  r_{22} &  0 \\ 
 r_{13}  &  0 &  r_{33}
\end{array}
\right], 
\label{expl}
\end{equation}
where entries labelled with a $c$ are complex and those labelled with an
$r$ are real. With these choices there are two complex entries 
in $m_{eff}$. From this equation we obtain:
\begin{equation}
I_{CP} =  {\rm Im} \ \left[ |m_{13}|^2 ( m_{12}^2  m_{22}^*  m_{11}^*) +
 |m_{12}|^2 ( m_{11}  m_{33}  {m_{13}^*}^2) +
 {m_{12}}^2 {m_{13}^*}^2  m_{22}^*  m_{33} \right].
\end{equation}
One may note that each one of the three terms contributing to $I_{CP}$
is rephasing invariant.  It is clear that $I_{CP}$ does
not vanish for the $m_{eff}$ of Eq.~(\ref{expl}).

It is well known that at low energies there are three CP 
violating phases in $U$, one of the Dirac type and 
two of the Majorana type. The question of finding the CP-odd WB
invariants that would be 
sufficient to control CP violation at low energies 
was first adressed in Ref.~\cite{Branco:1986gr}, and more recently in 
Ref.~\cite{Dreiner:2007yz}. In particular it was pointed out 
that requiring  the vanishing of 
the WB invariant of Eq.~(\ref{trr}) together with the two WB invariants:
\begin{eqnarray}
I_2 \equiv {\rm Im \ tr } \left[ h_l \; (m_{eff} \; m^*_{eff}) \;
( m_{eff} \; h^*_l \; m^*_{eff})\right],  \label{41} \\
I_3 \equiv {\rm {Tr}}\left[ \ (m^*_{eff}\  h_l\  m_{eff}\ ,\
h^*_l\right] ^3\ \label{42}
\end{eqnarray}
provides in general, necessary and sufficient conditions for low energy 
CP invariance \cite{Dreiner:2007yz}.
The invariant of Eq.~(\ref{42}), was first proposed in 
Ref. \cite{Branco:1998bw} where it was shown that it has the 
special feature of being sensitive to Majorana type CP violation even
in the limit of three exactly degenerate Majorana neutrinos.
Other relevant cases can be found in  Ref.~\cite{Branco:2005lp}.
The fact that, for the four zero texture of Eq.~(\ref{exe}),
none of the three WB invariants vanishes in general, shows that this
texture leads to both Dirac and Majorana-type CP violation at low energies.
The same applies to the other four zero textures.

So far, we have only considered leptonic CP violation at low energies.
Leptogenesis is a high energy phenomenon requiring CP violation.
In the unflavoured case the relevant phases are those in $m_D^\dagger m_D$ 
as shown in Eq.~(\ref{rmy}). In this case one may also write a set of
three independent WB invariants \cite{Branco:2001pq}
\begin{eqnarray}
I_4 \equiv {\rm Im Tr}[h_D H M_R^* h_D^* M_R], \label{i1} \\
I_5 \equiv {\rm Im Tr}[h_D H^2 M_R^* h_D^* M_R], \label {i2l}\\
I_6 \equiv {\rm Im Tr}[h_D H^2 M_R^* h_D^* M_R H],  \label{i3l}
\end{eqnarray}
where $h_D = {m_D}^\dagger m_D$ and $H = M_R^\dagger M_R$.
These three  would have to vanish if CP were to be conserved.  
The condition for the vanishing of $I_4$ was first given in Ref. 
\cite{Pilaftsis:1997jf}. 
The evaluation of these invariants, in the WB with diagonal $M_R$, 
shows that, in the case of a nondegenerate $D$ and assuming no
cancellations, they can all simultaneously vanish only if all
imaginary parts of  $({h_D}_{ij})^2$  are absent. 
Now it turns out that textures of category (ii) always have one zero 
off-diagonal entry in $h_{D}$ due to the orthogonality of two 
columns of $m_D$, but the other two off-diagonal elements are in
general nonzero. The same goes for those textures in category (i)
that also have two orthogonal columns. The remaining textures 
in category (i) have, in general,  three nonzero complex 
${h_D}_{ij}$ but their phases are constrained to be cyclic, i.e.,
${\rm Im} \ ({h_D}_{12}{h_D}_{31}{h_D}_{23}) = 0$.
The fact that not all imaginary parts of  $({h_D}_{ij})^2$  
vanish simultaneously in any of the four zero textures, shows that
they admit CP violation at high energies, relevant for leptogenesis.

\section{Implications from models with four zero textures}
\subsection{Low energy physics}

Let us start by summarising what is presently known about neutrino
masses and leptonic mixing.
We choose to parametrise the PMNS mixing matrix as
\cite{Yao:2006px}: 
\begin{equation}
U = \left( 
\begin{array}{ccc}
c_{12}c_{13} & s_{12}c_{13} & s_{13}e^{-i\delta } \\ 
-s_{12}c_{23}-c_{12}s_{23}s_{13}e^{i\delta } & \quad
c_{12}c_{23}-s_{12}s_{23}s_{13}e^{i\delta }\quad & s_{23}c_{13} \\ 
s_{12}s_{23}-c_{12}c_{23}s_{13}e^{i\delta } & 
-c_{12}s_{23}-s_{12}c_{23}s_{13}e^{i\delta } & c_{23}c_{13}
\end{array}
\right) \,\cdot P,  \label{pdg}
\end{equation}
where $c_{ij}\equiv \cos \theta _{ij}\ ,\ s_{ij}\equiv \sin \theta _{ij}\ $,
with all $\theta _{ij}$ in the first quadrant,  $\delta $ 
being a Dirac-type phase
and  $P=\mathrm{diag\ }(1,e^{i\alpha}, e^{i\beta})$ with 
$\alpha $ and $\beta$ denoting the phases associated with the
Majorana character of neutrinos.

The current experimental bounds on neutrino masses and leptonic
mixing are \cite{Yao:2006px}:
\begin{eqnarray}
\Delta m^2_{21} & = & 8.0 ^{+0.4}_{-0.3} \times 10^{-5}\  {\rm eV}^2, \\
\sin^2 (2 \theta_{12}) & = & 0.86 ^{+0.03}_{-0.04}, \\ 
|\Delta m^2_{32}| & = & (1.9 \ \  \mbox{to} \ \  3.0) \times 10^{-3}\  
{\rm eV}^2, \\
\sin ^2 ( 2 \theta_{23}) & > & 0.92, \\ 
\sin ^2  \theta_{13} &  < & 0.05, 
\end{eqnarray} 
with $\Delta m^2_{ij} \equiv m^2_j - m^2_i$. The allowed ranges
for the parameters listed above correspond to an impressive
degree of precision. The angle $\theta_{23}$ may be maximal
(i.e., $\pi / 4$). In contrast, maximal mixing for $\theta_{12}$
is already ruled out experimentally. At the moment there is only  
an experimental upper bound on the angle $ \theta_{13}$. 
A value of $ \theta_{13} $ close to the present bound 
would be good news for the prospects of detecting low energy leptonic 
CP violation, mediated through a Dirac-type
phase. The strength of the latter is given by:
\begin{equation}
{\cal J}_{CP} \equiv {\rm Im}\left[\,U_{11} U_{22}
U_{12}^\ast U_{21}^\ast\,\right] 
= \frac{1}{8} \sin(2\,\theta_{12}) \sin(2\,\theta_{13}) \sin(2\,\theta_{23})
\cos(\theta_{13})\sin \delta\,, \label{Jgen1}
\end{equation}
which would in this case be of order $10^{-2}$, for
$\sin \delta$ of order one.  A similar quantity defined in terms of the
elements of the Cabibbo Kobayashi Maskawa matrix is meaningful  
in the quark sector \cite{Jarlskog:1985ht} \cite{Bernabeu:1986fc},
and the corresponding value is of the order of $10^{-5}$.
It is not yet known
whether the ordering of the light neutrino masses is normal, i.e.
$m_1<m_2<m_3$, or inverted, i.e. $m_3<m_1<m_2$.
The scale of the neutrino
masses is not yet established. The spectrum may vary from a large  
hierarchy between the two lightest neutrino masses to three quasi-degenerate
masses. Examples of the possible extreme cases are: 
%\begin{eqnarray}
%{\rm (a)} \ \ \ \ \ m_1 \sim 0 \ \  (\mbox{or e.g.} \sim 10^{-6}
%\mbox{eV}),&\ \    
%m_2 \simeq 0.009 \ \mbox{eV}, \ \   m_3 \simeq 0.05 \ \mbox{eV}, \nonumber \\ 
%\mbox{implying a normal hierarchical spectrum.} \ \ \
%& \nonumber \\
%{\rm (b)} \ \ \ \ \ m_3 \sim 0 \ \  (\mbox{or e.g.} \sim 10^{-6}
%\mbox{eV}),& \ \  
%m_1 \simeq m_2  \simeq  0.05  \ \mbox{eV}, \nonumber \\
%\ \ \ \ \ \ \ \ \ \ \ \ \mbox{implying an inverted hierarchical spectrum.} &
%\nonumber \\ 
%{\rm (c)} \ \ \ \ \ m_1 \simeq 1 \ \mbox{eV}, \ \   m_2 \simeq 1 \ 
%\mbox{eV}, \hspace*{.5cm} &  m_3 \simeq 1 \ \mbox{eV}, 
%\nonumber \\ \mbox{implying a quasi-degenerate spectrum.} \ \ \ \ \ \ 
%& \nonumber
%\label{xxx}
%\end{eqnarray} 
\begin{eqnarray}
{\rm (a)} \ \ \ \ \ m_1 \sim 0 \ \  (\mbox{or e.g.} \sim 10^{-6} 
\mbox{eV}),\ \   
m_2 \simeq  0.009 
 \mbox{eV}, \ \   m_3 \simeq  0.05 \  \mbox{eV} \qquad \qquad \nonumber
\end{eqnarray}
corresponding to normal spectrum,  hierarchical, or else:
\begin{eqnarray}
{\rm (b)} \ \ \ \ \  m_3 \sim 0 \ \  (\mbox{or e.g.} \sim 10^{-6} 
\mbox{eV}), \ \ 
m_1 \simeq m_2  \simeq  0.05 \  \mbox{eV} \qquad \qquad \qquad \qquad \nonumber
\end{eqnarray}
corresponding to inverted spectrum, hierarchical, or else:
\begin{eqnarray}
{\rm (c)} \ \ \ \ \  m_1 \simeq 1 \mbox{eV}, \ \   m_2 \simeq 1 
\mbox{eV}, \ \   
m_3 \simeq 1 \mbox{eV} \qquad \qquad \qquad \qquad \qquad \qquad  \qquad 
\nonumber
\end{eqnarray}
corresponding to almost degeneracy.

As explained below, the conditions obtained in section 4,
are not all compatible with each of these scenarios. Finally, we note
that it is not yet established whether or not neutrinos
are Majorana particles and therefore at the moment there are
no restrictions on the Majorana phases $\alpha$, $\beta$ \\

The low energy implications of
patterns in category (i) which, as was already pointed out lead
to one off diagonal set of zeroes in the symmetric matrix $m_{eff}$
were studied in detail in Ref \cite{Merle:2006du}. 
The main conclusions in this 
paper are that no off-diagonal entry in  $m_{eff}$ can vanish
in the case of $\theta_{13}$ equal to zero. Implications of one zero 
in the first row of $m_{eff}$ do not differ much in the two possible 
such cases due to the approximate $\mu - \tau$ exchange symmetry 
\cite{Lam:2001fb}, \cite{Grimus:2001ex}, \cite{Ma:2002ce}. In this
case all values of $m_1$ are allowed from extreme hierarchy
to almost degeneracy; so are both possible orderings of neutrino
masses. For  $(m_{eff})_{23} = (m_{eff})_{32} =0 $
both normal hierarchy and inverted
hierarchy are excluded. \\

Let us consider category (ii) which implies the constraints of 
Eq.~(\ref{frm}). Taking into account Eq.~(\ref{eq7}) this equation 
can be written as:
\begin{equation}
\sum_{r<s} m_r m_s (U_{ir} U_{ks} - U_{is} U_{kr})
(U_{ir} U_{js} - U_{is} U_{jr}) = 0
\label{mrs}
\end{equation}
with i, j, k different from each other and no sum implied,
and the indices $r$, $s$ ranging from 1 to 3.

With the explicit parametrisation of Eq.(\ref{pdg}) we obtain simple
exact analytic relations for each of the classes in
category (ii).

For class (ii) (a) the exact form of this constraint is:
\begin{eqnarray}
& -  m_1 m_2 e^{2i\alpha}c_{23} s_{23} c_{13}^2 + \nonumber \\
& +  m_1 m_3 e^{2i\beta} [ c_{12}^2 c_{23} s_{23} + c_{12} s_{12} 
( c_{23}^2 -  s_{23}^2)  s_{13} e^{-i \delta} -  
s_{12}^2 c_{23} s_{23} s_{13}^2  e^{-2i \delta}] + \nonumber \\
& +  m_2 m_3  e^{2i(\alpha + \beta)} [  s_{12}^2 c_{23} s_{23} +
 c_{12} s_{12} ( s_{23}^2 -  c_{23}^2)  s_{13} e^{-i \delta}
- c_{12} c_{23} s_{23} s_{13}^2  e^{-2i \delta}] = 0 
\label{con1}
\end{eqnarray}
An interesting feature of this expression is the fact that 
all terms sensitive to Dirac type CP violation are doubly
suppressed since they are  multiplied, either by  $s_{13}^2$ or by 
the factor $( c_{23}^2 -  s_{23}^2)  s_{13}$, and it is already
known experimentally that $\theta_{13}$ corresponds to small 
or no mixing and  $\theta_{23}$ is maximal or close to maximal.
Therefore, this expression can be very well approximated by:
\begin{equation}
 -  m_1 m_2 e^{2i\alpha}\ c_{23} s_{23} c_{13}^2 +
m_1 m_3 e^{2i\beta}\  c_{12}^2 c_{23} s_{23} 
+  m_2 m_3  e^{2i(\alpha + \beta)}\    
s_{12}^2 c_{23} s_{23} = 0
\label{ap1}
\end{equation}

For class (ii) (b) we have the exact relation:
\begin{eqnarray}
 -  m_1 m_2 e^{2i\alpha}\  c_{23} c_{13} s_{13} e^{i \delta} +
m_1 m_3 e^{2i\beta}\  ( c_{12} s_{12} s_{23} c_{13} +
s_{12}^2 c_{23} c_{13} s_{13} e^{-i \delta}) + \nonumber \\
m_2 m_3  e^{2i(\alpha + \beta)}\  (- c_{12} s_{12} s_{23} c_{13} +
c_{12}^2 c_{23} c_{13} s_{13} e^{-i \delta}) = 0 
\label{con2}
\end{eqnarray}

Class (ii) (c) exactly verifies:
\begin{eqnarray}
 -  m_1 m_2 e^{2i\alpha}\  s_{23} c_{13} s_{13} e^{i \delta} +
m_1 m_3 e^{2i\beta}\  ( - c_{12} s_{12} c_{23} c_{13} +
s_{12}^2 s_{23} c_{13} s_{13} e^{-i \delta}) + \nonumber \\
m_2 m_3  e^{2i(\alpha + \beta)}\  ( c_{12} s_{12} c_{23} c_{13} +
c_{12}^2 s_{23} c_{13} s_{13} e^{-i \delta}) = 0 
\label{con3}
\end{eqnarray}
This equation can be  obtained from the previous one by interchanging
$s_{23}$ with $c_{23}$ and by changing the sign of the terms that
do not depend on the Dirac phase. 

It is clear from these expressions that the main features
of low energy physics coming out of these textures do not crucially
depend on the possible existence of CP violation.
In order to get a feeling of the main features of the 
implications of the constraints given by Eqs.~(\ref{con1}), (\ref{con2})
and (\ref{con3}), let us take as a first approximation the
Harrison, Perkins and Scott (HPS) mixing matrix \cite{Harrison:2002er}
\begin{equation}
U = \left[ 
\begin{array}{ccc}
 \frac{2}{\sqrt 6}  & \frac{1}{\sqrt 3} & 0 \\ 
 - \frac{1}{\sqrt 6}  &  \frac{1}{\sqrt 3} &  \frac{1}{\sqrt 2} \\ 
  - \frac{1}{\sqrt 6} & \frac{1}{\sqrt 3} & - \frac{1}{\sqrt 2}
\end{array}
\right],
\label{scott}
\end{equation}
which is consistent with present experimental data, and corresponds
to  $\theta_{23}$ maximal, $\theta_{13}$ zero and $c_{12} = 
2/\sqrt 6$ and no CP violation.
Obviously, a detailed analysis would require the variation
of $\theta_{13}$, as well as of $\theta_{12}$ and $\theta_{23}$, inside
their allowed ranges and also to take into consideration  
the possibility of CP violation.

Eqs.~(\ref{con1}) and (\ref{scott}) lead to 
\begin{equation}
\frac{1}{2} m_1 m_2 - \frac{1}{3} m_1 m_3 - \frac{1}{6} m_2 m_3 = 0.
\label{soma}
\end{equation}
and, as already explained, Eq.~(\ref{soma})
corresponds to ignoring terms with a 
double suppression. In the CP conserving limit, light neutrinos may have 
different CP parities \cite{Wolfenstein:1981rk}, therefore there are 
several possible ways of obtaining the necessary cancellations. 
Normal ordering with strong hierarchy is ruled out 
since for $m_1 \ll m_2$ there would be only one dominant term, the one 
in $m_2m_3$. Hierarchy in the
masses implies that the magnitude of term in $m_2m_3$ is close to $7 \times
10^{-5} \rm{eV}^2$.
The strongest allowed hierarchy
consistent with the above constraint
favours the larger $\theta_{13}$ values; a numerical example
obtained for maximal  $\theta_{23}$ and the central value of
$\theta_{12}$, with cancellations already of order $10^{-10} \rm{eV}^2$
is:
\begin{eqnarray}
m_1 = 0.00333071221  \rm{eV}, &\ m_2 = - 0.00954429898  \rm{eV}, \qquad
 m_3 = 0.0501108132 \rm{eV}, \nonumber \\
s_{13} = 0.198669  & \nonumber
\end{eqnarray}
Obviously the number of significant digits in the above numerical 
result is meaningless from the experimental point of view and is
only given to be consistent with the degree of cancellation 
claimed above.
Here $m_2$ is no more than a factor of three higher than $m_1$, 
which corresponds to a weak normal hierarchy.
As $m_1$ decreases, cancellations cease to occur and the difference
tends to the value of the dominant term which is the term 
in $m_2m_3$. The situation would change for larger values of 
$\theta_{13}$ which are already ruled out. Likewise for instance, 
for a small solar angle, ($\theta_{12}$), already excluded, 
as was pointed out in \cite{Branco:2002xf},
where a particular example of a texture of this class was considered,
since in this case the term in  $m_2m_3$ present  in Eq.~(\ref{ap1})
would be suppressed by $s^2_{12}$.
Strong inverse hierarchy is also ruled out since it would leave the 
dominant term in $m_1m_2$ without the possibility of cancellation.
Quasi-degeneracy can be accommodated within the present range
of experimental values for the mixing angles. \\

Case (b) in category (ii) obeys the constraint of Eq.~(\ref{con2}).  
The coefficient of $m_1m_2$ is zero  
for the HPS matrix and  in this case we are left with 
\begin{equation}
- \frac{1}{3} m_1 m_3 + \frac{1}{3} m_2 m_3  = 0.
\label{nun}
\end{equation}
For nonzero $\theta_{13}$  the term in $m_1m_2$ is suppressed but 
cannot be discarded for $m_3 \ll m_1, m_2$, i.e., inverse hierarchy. 
In the case of inverse hierarchy the necessary cancellation of the 
three terms may occur.  Almost degeneracy can also be accommodated
provided $\theta_{13}$ is different from zero. For  $\theta_{13} = 0$ the
coefficient of  $m_1m_3$ would be exactly equal to the 
coefficient of $m_2m_3$ and this relation could not be verified,
since it would imply $m_1 = m_2$.

Case (c) in category (ii) is very similar to case (b). 
The resulting equation for the HPS matrix coincides 
with Eq. (\ref{nun}) and the conclusions are the same as in case (b),
category (ii). \\ 

All cases in category (ii) are thus incompatible with a strong
hierarchy and normal ordering, i.e. for $m_1 \ll m_2 $. \\

\subsection{Relating leptogenesis to low energy physics}

Zero textures in $m_D$ allow one to relate the matrix
$R$, relevant to leptogenesis,
to the light neutrino masses and low energy leptonic mixing. In fact,  
it is clear from Eq. (5) that each zero in $m_D$ leads to an orthogonality
condition between one column of
the matrix $R$ and one row of the matrix $U \sqrt{d}$ of the form: 
\begin{equation}
(m_{D})_{ij}=0~ \Rightarrow ~ (U)_{ik}\sqrt{d}_{kk}R_{kj}=0.  
\label{orto}
\end{equation}
It was already pointed out \cite{Branco:2005jr}
that the connection between leptogenesis and low energy physics 
could be easily established in a particular case that falls into category
(ii), since in this case one can fully express the matrix
$R$ in terms of light neutrino masses and low energy leptonic mixing . 
The same is true for all other cases in category (ii) as well as for
the cases that fall into
category (i) as shown below. The example given in Ref. \cite{Branco:2005jr} 
can be generalised in the following way. \\

In category (i) there is always in  $m_D$  one column with two zeros
and two columns with one zero each. Let $l$ be the column with two
zeros and $a$ and $b$ the columns with  one zero only. In this case we
can write
\begin{eqnarray}
\left( \vec{R_l} \right)_i = \left( \varepsilon_{ijk} (U)_{pj}
\sqrt{{m}_{j}} \; 
(U)_{qk}\sqrt{{m}_{k}} \right) \; \frac{1}{N_l},
\label{rli}  \\
\left( \vec{R_a} \right)_i = \left( \varepsilon_{ijk} (U)_{rj}
\sqrt{{m}_{j}} \; 
\left( R_l \right)_k  \right) \; \frac{1}{N_a},
\label{outros} \\
\left( \vec{R_b} \right)_i = \left( \varepsilon_{ijk} (U)_{sj}
\sqrt{{m}_{j}} \; 
\left( R_l \right)_k  \right) \; \frac{1}{N_b},
\label{mais}
\end{eqnarray}
where p and q are the rows with zeros in the column $l$.  Moreover, r
and s are the rows 
where the zeros are in columns $a$ and $b$ respectively.  
The $\vec{R_i}$ are the columns of the matrix $R$ 
and the $N_{i}$ are complex normalization factors, with
phases such that $\vec{R_i}^2 = 1$. It is easy to show that 
the columns $\vec{R_a}$ and $\vec{R_b}$ are indeed orthogonal to
each other by using the constraint ${m_{eff}}_{rs} =0$
valid for each case in category (i). 

In Ref.~\cite{Branco:2002xf} the relation between leptogenesis and $CP$
violation at low energies in two cases falling into category (i),
were analysed in detail in the case of hierarchical heavy
Majorana neutrinos and also in the case of two fold quasi-degeneracy
of the heavy neutrinos. \\

In category (ii) there is in  $m_D$ always one column without zeros and
each 
of the other two columns has two zeros. Now let $l$ be the column without 
zeros and $a$ and $b$ the columns with two zeros. In this case we 
can write:
\begin{eqnarray}
\left( \vec{R_a} \right)_i = \left( \varepsilon_{ijk} (U)_{pj}
\sqrt{{m}_{j}} \; 
(U)_{qk}\sqrt{{m}_{k}} \right) \; \frac{1}{N_a}, \\
\left( \vec{R_b} \right)_i = \left( \varepsilon_{ijk} (U)_{rj}
\sqrt{{m}_{j}} \; 
(U)_{sk}\sqrt{{m}_{k}} \right) \; \frac{1}{N_b}, \\
\left( \vec{R_l} \right)_i = \varepsilon_{ijk}
\left( \vec{R_a} \right)_j \;  \left( \vec{R_b} \right)_k. 
\label{gggg}
\end{eqnarray} 
Here p and q are the rows with zeros in column $a$. Furthermore, 
r and s are the rows with zeros in column $b$, while $N_i$ denote
normalization factors. It is easy to show that  
$\vec{R_a}$ and $\vec{R_b}$ are indeed orthogonal to
each other for each case in category (ii) by using Eq. (19).  \\

All four zero textures analysed in this paper
allow one to completely specify the matrix $R$, in terms
of light neutrino masses and the elements of the PMNS matrix. 
It is clear that $R$ can only be complex
if there is CP violation at low energies.

\section{Summary and Conclusions}

We have made a systematic study of all allowed four zero textures
in the neutrino Dirac mass matrix $m_D$, in the framework of Type I 
seesaw mechanism, without vanishing neutrino masses. In order for 
this study to be meaningful, one has to choose a specific
weak basis (WB). Without loss of generality, we have chosen to work
in the WB where the charged lepton and the righthanded neutrino mass
matrices are both diagonal, real. Assuming that no neutrino mass
vanishes and taking into account the experimental evidence
that no leptonic family decouples from the other two, we have shown
that four is the maximal number of zeros allowed in $m_D$. We have found
the following remarkable result: the allowed four zero textures 
in the neutrino Yukawa coupling matrices automatically lead to one
of two patterns in $m_\nu$. Either the latter has only one pair of 
symmetic off-diagonal zeros or else it has no zero element but
a vanishing subdeterminant condition. In the derivation of this
result, we have implicitly assumed the absence of any fine-tuning
between the parameters of $m_D$ and those of $M_R$ which would lead to
special cancellations.

Our analysis also applies to scenarios where instead of zeroes, 
one has a set of extremely suppressed entries \cite{Plentinger:2007px}, 
as one often 
encounters in the Froggatt-Nielsen approach. Of course,
renormalisation group effects, especially for quasi-degenerate 
and inverted hierarchical neutrinos  \cite{Dighe:2007nh} 
will change at least some 
of the zeroes in $m_D$ into small entries.

We have also explored the phenomenological consequences of the
above mentioned textures. In particular we have shown that they 
lead to a close
connection between leptogenesis and low energy measurables such as
neutrino masses and mixing angles. The establishment of such a
connection in the leptonic sector between physics at low and very
high energies is an important goal and provides an additional motivation 
for considering texture zeroes in the leptonic sector.

\section*{Acknowledgements}
This work was partially supported by Funda\c c\~ ao para a 
Ci\^ encia e a  Tecnologia (FCT, Portugal) through the projects
POCTI/FNU/44409/2002, PDCT/FP/63914/2005, PDCT/FP/63912/2005 and
CFTP-FCT UNIT 777 which are partially funded through POCTI 
(FEDER). The work of D.E.C. is presently supported by a CFTP-FCT UNIT 777
fellowship. The work of G.C.B. was supported by the Alexander 
von Humboldt Foundation through a Humboldt Research Award. 
The work of P.R. has been supported in part by DAE (BRNS).
G.C.B. would
like to thank Andrzej J. Buras for the kind hospitality at TUM. P.R. 
acknowledges the generous hospitality of CFTP (IST), Lisboa, where this
investigation started. G.C.B. and M.N.R. are grateful for the warm 
hospitality of the CERN Physics Department (PH) Theory Division (TH) where the 
work was finalised.

\end{document}